\documentclass[fleqn,usenatbib]{mnras}

\usepackage[T1]{fontenc}
\usepackage{newtxtext,newtxmath}

\usepackage{graphicx}
\usepackage{amsmath}
\usepackage[nopatch]{microtype}
\usepackage{booktabs}

\usepackage{float}
\usepackage{placeins}
\usepackage{hyperref}
\usepackage{caption}
\usepackage{orcidlink}
\usepackage{longtable}

\title{Search for intranight optical variability in a large sample of intermediate-mass black holes}

\author[Sharma et al. ]{
Himanshu Sharma \orcidlink{0009-0006-8185-7322}$^{1}$ \thanks{E-mail: himanshu4gya@gmail.com},
Gopal-Krishna$^{2}$,
Vibhore Negi \orcidlink{0000-0001-5824-1040}$^{3}$ \thanks{E-mail: vibhore.negi18@gmail.com},
Hum Chand\orcidlink{0000-0002-3163-4941}$^{1}$,
Krishan Chand \orcidlink{0000-0002-6789-1624}$^{1}$,
\newauthor
Anshul Kumar Sharma \orcidlink{0009-0009-5248-6907}$^{1}$
\\
$^{1}$Department of Physics and Astronomical Science, Central University of Himachal Pradesh, Dharamshala, 176215, India\\
$^{2}$UM-DAE Centre for Excellence in Basic Sciences, Vidyanagari, Mumbai-400098, India\\
$^{3}$Kavli Institute for Astronomy and Astrophysics, Peking University, Beijing, 100871, People's Republic of China \\
}

\date{Accepted XXX. Received YYY; in original form ZZZ}
\pubyear{2026}

\begin{document}

\label{firstpage}
\pagerange{\pageref{firstpage}--\pageref{lastpage}}
\maketitle

% Abstract of the paper

\begin{abstract}
Using the extensive archival database of the Zwicky Transient Facility (ZTF) survey, we have investigated intranight optical variability (INOV) of a large sample of low-mass Active Galactic Nuclei (LMAGN) powered by Intermediate Mass Black Holes (IMBH). After applying a sequence of well motivated selection filters, we have built an unbiased, representative sample of 62 IMBH for which $r$-band intranight photometric sequences are available in the ZTF database, yielding 167 sessions of minimum 2 hour duration. By performing aperture photometry on the sequence of exposures in each session, we derived differential light-curves (DLCs) of the target LMAGN, relative to two carefully selected non-varying comparison stars monitored simultaneously with the LMAGN in the same exposure sequence. Application of the widely used test based on $F$-statistics to all these DLCs has revealed no case of statstically significant INOV. We discuss this result for the present large sample of LMAGN in the context of an earlier report of INOV detections in another, albeit much smaller and differently selected sample of LMAGN, which had been monitored in targeted observations.
\end{abstract}


\section{Introduction}
\label{sec:intro}
Black hole phenomenology, mostly probed through the electromagnetic window, has remained largely focused at two opposite ends of the mass spectrum. Such black holes are thus classified either as (i) supermassive black holes (SMBHs), for which the mass ranges between $10^{7}\,M_{\odot}$ - $10^{10}\,M_{\odot}$, or (ii) stellar-mass black holes, having masses of order $10\,M_{\odot}$ (e.g., \citealt{karet_20217}; \citealt{Marti2017}; \citealt{blandford2019}). Much uncertainty exists, however, about the so-called intermediate-mass black holes (IMBHs), thought to populate the mass range $10^{3}$--$10^{6}\,M_{\odot}$. Currently, the most secure example of an IMBH (near the high end of this mass range) is Sgr A$^{*}$ located at the centre of our Galaxy, for which a robust mass estimate of $\sim4\times10^{6}\,M_{\odot}$ is available, based on kinematical measurements of the stars orbiting it \citep{ghez_2008, genzel_2010}. At the same time, mass estimates for a large number of IMBH candidates are now available, based on single-epoch optical spectra, employing the luminosity and width of a prominent broad emission line and leveraging the luminosity--line width scaling relations established for SMBHs \citep{Kaspi_2000, greene_ho_2004, vestergard_peterson_2006}. While being aware of the caveats involved in this approach, large catalogues containing several hundred IMBH candidates have been published (e.g., \citealt{dong_2012}; \citealt{liu_2018}; \citealt{chilingarian_2018}).

Nearly all massive galaxies are believed to harbour a SMBH at the nucleus \citep{Kormendy1995, richstone_1998}. A key development in this context is the fairly tight correlation observed between the SMBH mass and the stellar mass of the host galaxy \citep{Magorrian1998, haring_2004, mconnell_2013, Kormendy2013, Reines2015,li_2023}. This correlation clearly points to a coordinated growth of the two entities, probably mediated by kinematic feedback from thermal/nonthermal outflows injected by the SMBH into the surrounding gaseous medium associated with the host galaxy \citep[e.g.,][]{Fabian2012, Beckmann2017, Morganti2017, Mukherjee2018}. Extrapolation of this correlation suggests that the nuclei of low-mass (dwarf) galaxies having stellar masses between $\sim10^{8}\,M_{\odot}$ and $10^{10}\,M_{\odot}$, are the most likely sites for IMBHs, although this prognosis remains uncertain in some numerical simulations \citep{Barai2019, Dubois2021, ArjonaGalvez2024}.

A major challenge in discerning any AGN activity associated with IMBHs is that the expected signatures can be mimicked by intense star formation occurring in the nuclear region of the host galaxy. This has given impetus to the use of optical broadband flux variability as a tool for identifying AGN activity in such systems \citep{Baldassare2018,Elmer2020, Kimura2020, Ward2022}. Using this approach, it has recently been inferred that the incidence of AGN activity in dwarf galaxies may be comparable to that found in massive galaxies, although this conclusion is still based on a relatively small sample \citep{Kaviraj2025}. Compared to the long-term optical variability, an even stronger diagnostic of relativistic jet activity is provided by intranight optical variability (e.g., \citealt{GopalKrishnaWiita2018}). Motivated by this possibility, an INOV monitoring campaign was carried out by some of us, for a sample of 12 radio- and X-ray-detected low-mass AGNs (LMAGNs) supposed to be powered by an IMBH at the nucleus. That study yielded an estimate of INOV duty cycle of $\sim22\%$ for variability amplitudes $\psi>3\%$, which was, albeit, considered unexpectedly high for this class of objects \citep[][hereafter Paper~I]{gopal_krishna}. Given the small size of that sample and its pre-selection based on confirmed detection in both radio and X-ray bands, it was considered worthwhile to carry out an independent INOV search using a much larger sample of LMAGNs (IMBH candidates) which is also unbiased with respect to radio/X-ray detection. Such a study has now become feasible owing to the extensive intranight optical monitoring data available from the Zwicky Transient Facility (ZTF, see \citealt{ztf}).

In this paper, we analyse the archival ZTF data to investigate the incidence of INOV in a large, well-defined sample of LMAGN powered by IMBH. Section~\ref{sec:sample} describes the assembling of such a sample and the selection of the ZTF monitoring sessions analysed in this work. The data reduction, involving aperture photometry, construction of differential light curves (DLCs), and their statistical analysis, together with the results of the INOV search, are presented in Section~\ref{sec:analysis_results}, followed by the discussion and conclusions in Sections~\ref{sec:discussion} and \ref{sec:conclusions}, respectively.

% The data reduction involving aperture photometry, derivation of differential light curves (DLCs), and their statistical analysis are presented in Section~\ref{sec:analysis_results}. The results are presented in Section~\ref{sec:results}, followed by discussion and conclusion in Section~\ref{sec:discussion} and Section~\ref{sec:conclusions}, respectively.

%%%%%%%%%%%%%%%%%%%%%%%%%%%%%%%%%%%%%%%%%%%%%%%%%%%%%%%%%%%%%%%%%%%%%%%%%%%%%%%%%%%%%%%%%%%%%%%%%%%%%%%%%%%%%%%%%%%%%%%%%%%%%%%%%%%%%%%%%%%%%%%%%%%%%%%%%%%%%%%%%%%%%%%%%

\section{Sample Selection}
\label{sec:sample}

\subsection{The Parent sample of low-mass AGN}

Our parent sample is assembled by combining the optically selected catalogues of LMAGN compiled by \citet{dong_2012} and \citet{liu_2018}. \citet{dong_2012} identified 309 Type-1 AGNs from the SDSS DR4 spectroscopic database \citep{sdss_dr4}, selected through the detection of the broad H$\alpha$ emission line. Black hole masses were estimated using the width and luminosity of the broad H$\alpha$ line together with the single-epoch virial mass estimator, resulting in masses in the range
$8\times10^{4} < M_{\rm BH} < 2\times10^{6}\,M_\odot$.

Using the same methodology, \citet{liu_2018} identified an additional 204 LMAGN from SDSS DR7 spectra \citep{sdss_dr7}, with black hole masses estimated to lie in the range
$1\times10^{5} < M_{\rm BH} < 2\times10^{6}\,M_\odot$.

\begin{figure*}
    \centering
    \includegraphics[width=.90\textwidth]{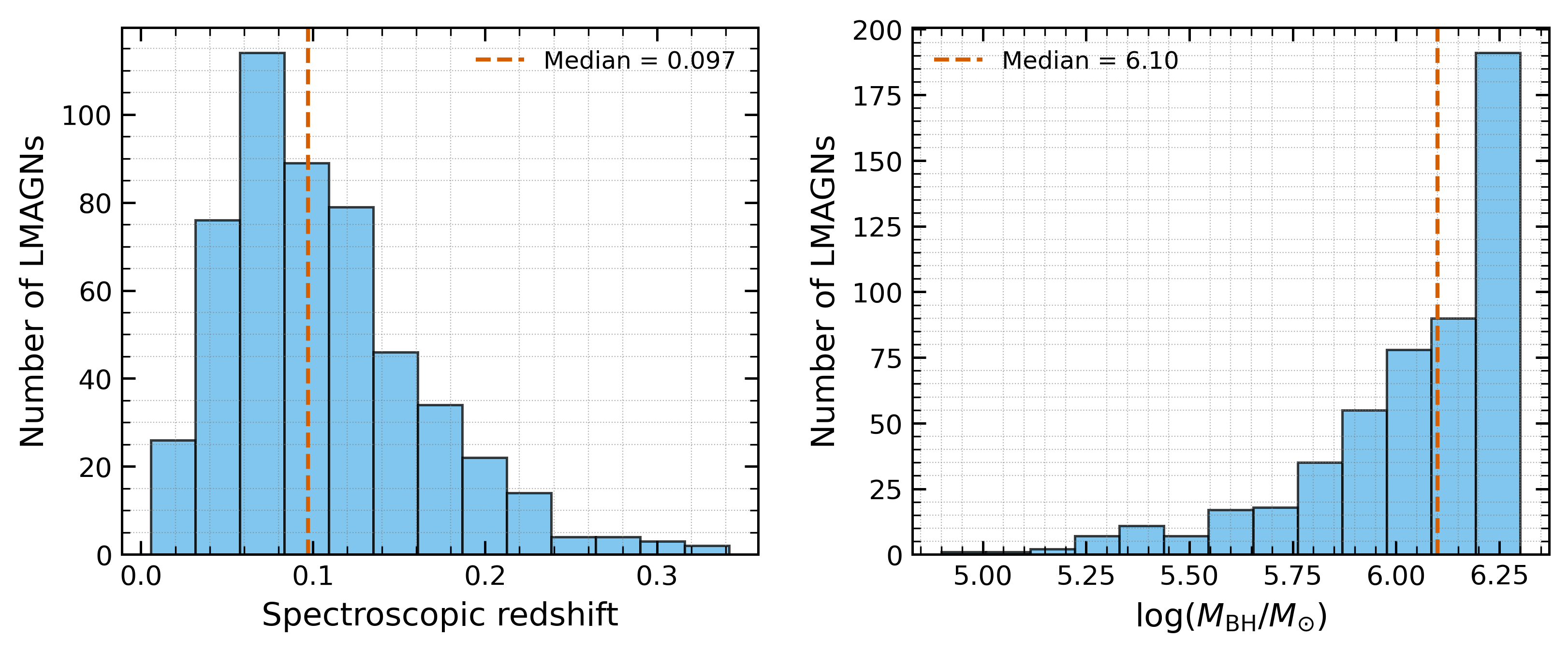}
    \caption{Distributions of the spectroscopic redshift (left) and black hole mass (right) for the parent sample containing 513 low-mass AGNs. The dashed vertical lines indicate the median values of the corresponding distributions.}
    % \caption{Distributions of (top) Spectroscopic redshift and (bottom) black hole mass for the parent sample of 513 low-mass AGNs.}
    \label{bh_prop}
\end{figure*}

The combined catalogue thus contains 513 LMAGNs with redshifts $z<0.35$, a median redshift of 0.097 and a median $M_{BH}$ of $\log(M_{\rm BH}/M_\odot)=6.1$. The distributions of $z$ and $M_{BH}$ for this parent sample are shown in Fig.~\ref{bh_prop}.

\subsection{Selection of ZTF counterparts and the intranight light curves}
\label{subsec:ztf_selection}

We searched for archival $r$-band light curves of the 513 LMAGN in the ZTF DR21 release \citep{ztf}. A series of objective quality cuts was then applied to shortlist observing sessions suitable for investigating the INOV.

\begin{enumerate}

\item \textbf{ZTF positional matching:}
A positional search within a radius of 5 arcsec yielded ZTF counterparts for 498 of the 513 sources.

\item \textbf{Photometric quality threshold:}
Only measurements with \texttt{CATFLAG}=0 were retained, in order to exclude less reliable photometry. This excluded one source, leaving 497 sources for further analysis.

\item \textbf{Selection of a unique observation ID and minimum intranight measurements:}
Since ZTF archives observations obtained from different fields and CCD quadrants as separate light curves, each identified by a unique observation ID, for each source we retained only the observation ID containing the largest number of measurements \citep{negi_2023}. This avoids combining photometric measurements potentially obtained with different instrumental configurations. For the selected observation ID, measurements obtained on the same night were then grouped into individual observing sessions. Only sessions containing at least 20 intranight measurements (i.e. data points) were retained for further consideration. This yielded a total of 289 observing sessions pertaining to 84 LMAGN.
% For each source, individual observing sessions were identified by grouping measurements obtained during the same night. Only sessions containing at least 20 data points were retained, yielding 289 observing sessions corresponding to 84 sources.

\item \textbf{Duration and cadence cuts:}
To ensure a reasonable temporal coverage, we mandated each qualifying session to have (i) an overall duration exceeding 2 hours and (ii) no time gap larger than 0.5 hours between consecutive measurements. These criteria reduced the sample to 191 observing sessions pertaining to 69 LMAGNs, and these data were subjected to aperture-photometry and subsequent analysis (Section~\ref{sec:analysis_results}).

% sources. Eleven of these sessions had to be discarded because the target AGN was contaminated by a neighbouring object and thus our final sample consists of 180 observing sessions pertaining to 66 low-mass AGN.

\end{enumerate}

% The step-wise selection procedure is summarised in Table~\ref{tab:sample_summary}.

% \begin{figure}
%     \centering
%         \includegraphics[width=0.48\textwidth]{duration_hist_192_srcs.png}
%         \caption{Distribution of time duration for the final 84 INOV sessions in the observed frame, with a median value of 3.87 hr.}
%         \label{mbh_hist}
% \end{figure}

\begin{table*}
\centering
\caption{Summary of the sequence of selection cuts applied to construct the final INOV sample of 62 low-mass AGN (LMAGN).}
\label{tab:sample_summary}
\begin{tabular}{clcc}
\toprule
Step & Selection filter applied & LMAGN Left & Sessions Left \\ % & Notes \\
\midrule
0 & Parent low-mass AGN sample (\citet{dong_2012} + \citet{liu_2018}) & 513 & -- \\ % & $z<0.35$ \\
1 & The ZTF positional matching ($<5$ arcsec) & 498 & -- \\ % & Matched in ZTF DR21 \\
2 & Photometric quality check application: CATFLAG = 0 & 497 & -- \\ % & Removed 1 source \\
% 3 & Selecting only the ZTF obsID with maximum number of measurements & 497 & -- \\ % & Avoid mixing fields/CCDs \\
% 4 & At least 20 measurements (data points) in the session & 84 & 289 \\ % & Initial session extraction \\
3 & ZTF obsID with maximum measurements + $\geq20$ points per session & 84 & 289 \\ % & Initial session extraction \\
4 & Total session duration $>2$ hr and any gap $<0.5$ hr & 69 & 191 \\ % & Cadence-quality cut \\
5 & Flagging segament(s) with unsteady PSF and then discarding the session if the remaining DLC is shorter than 2 hr & 66 & 179 \\ % &  \\
6 & Excluding sessions where the LMAGN image is affected by neighbouring-source contamination & 62 & 167 \\ % & Visual Inspection \\
% 7 & Mean photometric error $<0.03$ mag & 23 (1bin) 55(3bin) & 72 (1bin) 153(3bin) & Mean error cut \\
% 8 & Remove poorly sampled source J160656... & ... & ... & 11 pts + high noise \\
% 6 & Seeing–magnitude correlation: $\rho > -\rho_{\rm crit}$ & 39 & 67 & Ensures host–seeing stability \\
% 7 & Radio-quiet cut using $R = F_{1.4GHz}/F_B$ & 36 & 63 & Remove 1 source (3 sessions) \\
\midrule
 & \textbf{The final sample of INOV sessions (after applying all the 6 selection filters)} & \textbf{62} & \textbf{167} \\ % & Used for F-test INOV analysis \\
\bottomrule
\end{tabular}
\end{table*}

%%%%%%%%%%%%%%%%%%%%%%%%%%%%%%%%%%%%%%%%%%%%%%%%%%%%%%%%%%%%%%%%%%%%%%%%%%%%%%%%%%%%%%%%%%%%%%%%%%%%%%%%%%%%%%%%%%%%%%%%%%%%%%%%%%%%%%%%%%%%%%%%%%%%%%%%%%%%%%%%%%%%%%%%%
\section{Data Analysis and Results}
\label{sec:analysis_results}

\subsection{Aperture-photometry}
\label{subsec:aperture}

The ZTF archive provides PSF-based photometry for all the detected sources. However, as our LMAGNs are nearby ($z<0.35$), the host galaxy can contribute significantly to the measured peak flux, and variations in the seeing disk, i.e. `Point Spread Function (PSF)', as well as PSF-fitting/model systematics associated with the spatially extended host-galaxy emission, may introduce spurious variability into the ZTF PSF-based photometry \citep{cellone_2000, Arevalo2026}. To minimise this possibility, we carried out fixed aperture photometric measurements using the DAOPHOT package \citep{daophot} on the ZTF science images corresponding to our entire set of intranight observing sessions.

For each of the intranight sessions, ZTF science image was downloaded and the FWHM for each object detected in that image was fitted with a two-dimensional Gaussian profile. The mean FWHM of the detected objects was then computed for that frame, and the median of these values taken over all the frames in that observing session was adopted as the representative PSF for that session. A constant aperture with a radius equal to twice this `session PSF' was then used for aperture photometry of both the target LMAGN and the comparison stars (see below) throughout that intranight session. 
In any case, to stay on the conservative side, we have confined the INOV search to only those segments of the photometric sequence (session) over which the PSF was found to remain steady.

Instrumental magnitude of a given object in the frame was measured using DAOPHOT, estimating the local sky background through an annular sky aperture surrounding the object. These instrumental magnitudes were subsequently used to construct differential light curves (DLCs) of the object with respect to the chosen two comparison stars, as described in the following subsection.

\subsection{Selection of the comparison stars and construction of the differential light curves}
\label{subsec:comparison_stars}

For each target LMAGN, a set of candidate comparison stars was initially shortlisted through visual inspection of the ZTF images. Out of this set, two comparison stars called (S1 and S2) were selected after satisfying the following criteria: (i) Image is not saturated, (ii) Brightness differs from that of the target LMAGN by less than one magnitude, (iii) Separation from the LMAGN is sufficiently large in order to avoid contamination from it, or its host galaxy, and (iv) Sufficient isolation from neighbouring sources in order to minimise blending effects. The same pair of comparison stars was used throughout a given observing session.

Using the instrumental magnitude measurements from our aperture photometry, three differential light curves (DLCs) were constructed for each observing session, namely Q--S1, Q--S2, and S1--S2, where Q denotes the target LMAGN. The Q--S1 and Q--S2 DLCs were used to assess the presence/absence of INOV of the LMAGN, while the S1--S2 DLC served as a control to be certain of the photometric stability of the comparison stars during the observing session. Steadiness of this `star-star' DLC is a pre requisite for genuineness of any variations detected in the two DLCs of the target LMAGN. 

\subsection{Quality assessment of the differential light curves}
\label{subsec:quality}

The differential light curves (DLCs) obtained from the aperture photometry were visually examined in order to assess their suitability for INOV search. Since our sample consists of nearby LMAGNs with appreciable, often dominant contribution coming from the host-galaxy, PSF changes during an observing session can alter the amount of host-galaxy light (relative to the AGN) enclosed within the fixed aperture adopted for that session (see Section~\ref{subsec:aperture}), thereby introducing spurious INOV into the DLCs of the target LMAGN.

\begin{figure}
    \centering
        \includegraphics[width=.5\textwidth]{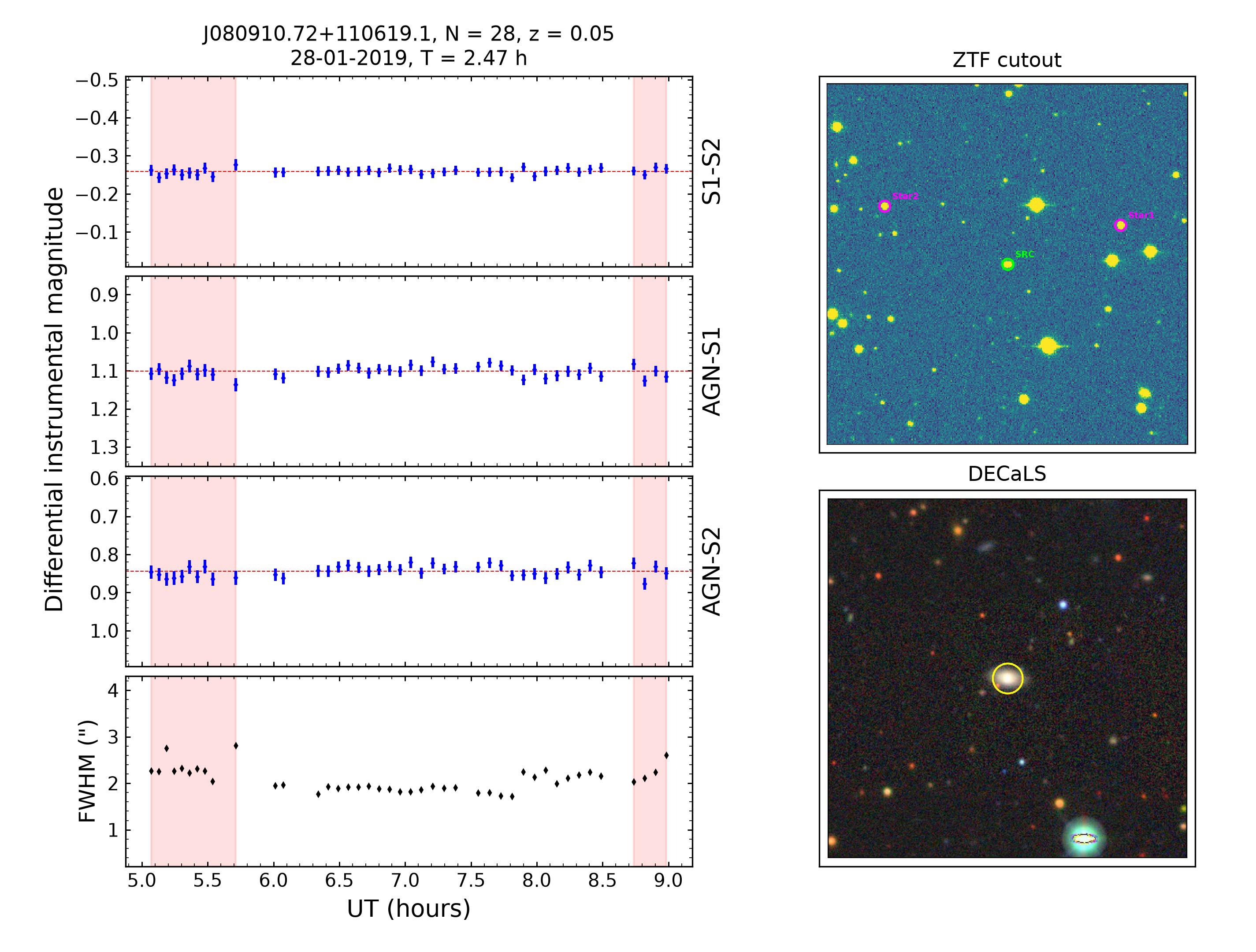}
        \caption{A representative example illustrating the outcome of the analysis carried out in this work. The upper-right and lower-right panels show the ZTF and DECaLS images, respectively, with the LMAGN and the selected pair of comparison stars marked. The ZTF and DECaLS image cutouts have sizes of $5\,\mathrm{arcmin} \times 5\,\mathrm{arcmin}$ and $75\,\mathrm{arcsec} \times 75\,\mathrm{arcsec}$, respectively. The left panels show the seeing disk (PSF) variation together with the differential light curves constructed from our aperture photometry. The red shaded regions indicate intervals affected by unstable seeing, hence excluded from the analysis, including the application of the F-test, but are retained in the plots for completeness. Similar figures were generated and analysed for all the 167 observing sessions.} 
        \label{fig:representative_dlc}
\end{figure}

To guard against this possibility, we inspected the PSF curve for each session in order to identify the time segment over which the PSF remained fairly steady (i.e., no systematic gradient seen). The remaining time segments which we have conservatively chosen to exclude from the analysis, are shown as shaded regions (Fig.~\ref{fig:representative_dlc}). If, after this trimming, the duration of the DLC became shorter than 2 hr, we dropped that session from further consideration. As a result, 12 observing sessions had to be discarded. A further 12 sessions for which the image of the target LMAGN was seen to be affected by contamination from nearby objects were also discarded, leaving a final sample of 167 observing sessions corresponding to 62 LMAGNs, for INOV search based on statistical analysis. The entire step-wise selection procedure is summarised in Table~\ref{tab:sample_summary}.
Note also that, for a few DLCs with unusually dense sampling,  additional smoothing was carried out by 3-point box averaging. Particulars of the 167 qualifying sessions are given in Table~\ref{tab:inov_results}.
% A representative example is shown in Fig.~\ref{fig:representative_dlc}.

% \section{Results}
% \label{sec:results}

\subsection{Photometric sensitivity of the differential light curves (DLCs)}
\label{subsec:photometric_sensitivity}

% To begin with, we would like to put in perspective the sensitivity of the archival ZTF monitoring data used here. For this, we shall compare the typical photometric precision of the DLCs with that achieved in the DFOT-based monitoring of another IMBH sample which was reported in Paper~I. For each of the two AGN DLCs in a given session, we define a photometric noise parameter (PNP), based on the rms photometric error of its individual data points. The rms uncertainties returned by DAOPHOT for each measurement were multiplied by the error-scaling factor $\eta=1.54$ in order to account for the systematic underestimation of the formal photometric errors by DAOPHOT \citep{Goyal2013}. The PNP for a DLC is thus defined as

To begin with, we would like to put in perspective the sensitivity of the archival ZTF monitoring data used here. For this, we shall compare the typical photometric precision of the DLCs with that achieved in the monitoring of another IMBH sample using the 1.3-m Devasthal Fast Optical Telescope (DFOT), as reported in Paper~I. For each of the two AGN DLCs in a given session, we define a photometric noise parameter (PNP), based on the rms photometric error of its individual data points. Since the formal photometric uncertainties returned by DAOPHOT are known to be systematically underestimated, they were scaled up by a factor ($\eta$), for which we adopt ($\eta=1.54$), following \citet{Goyal2013eta}. The PNP for a DLC is thus defined as

\begin{equation}
{\rm PNP}
=
\sqrt{\eta^{2}
\left\langle \sigma_{i,\rm err}^{2}\right\rangle}
=
\eta
\sqrt{
\frac{1}{N}
\sum_{i=1}^{N}
\sigma_{i,\rm err}^{2}
},
\label{eq:pnp}
\end{equation}

where $\sigma_{i,\rm err}$ is the rms photometric error of the $i$-th data point as returned by the DAOPHOT algorithm, $N$ is the number of data points in the DLC, and the angular brackets denote averaging over the N data points. Since two AGN DLCs, Q--S1 and Q--S2, were derived for each observing session, the average of their PNP values was taken as the measure of the photometric noise level for that session.

\begin{figure}
    \centering
    \includegraphics[width=0.48\textwidth]{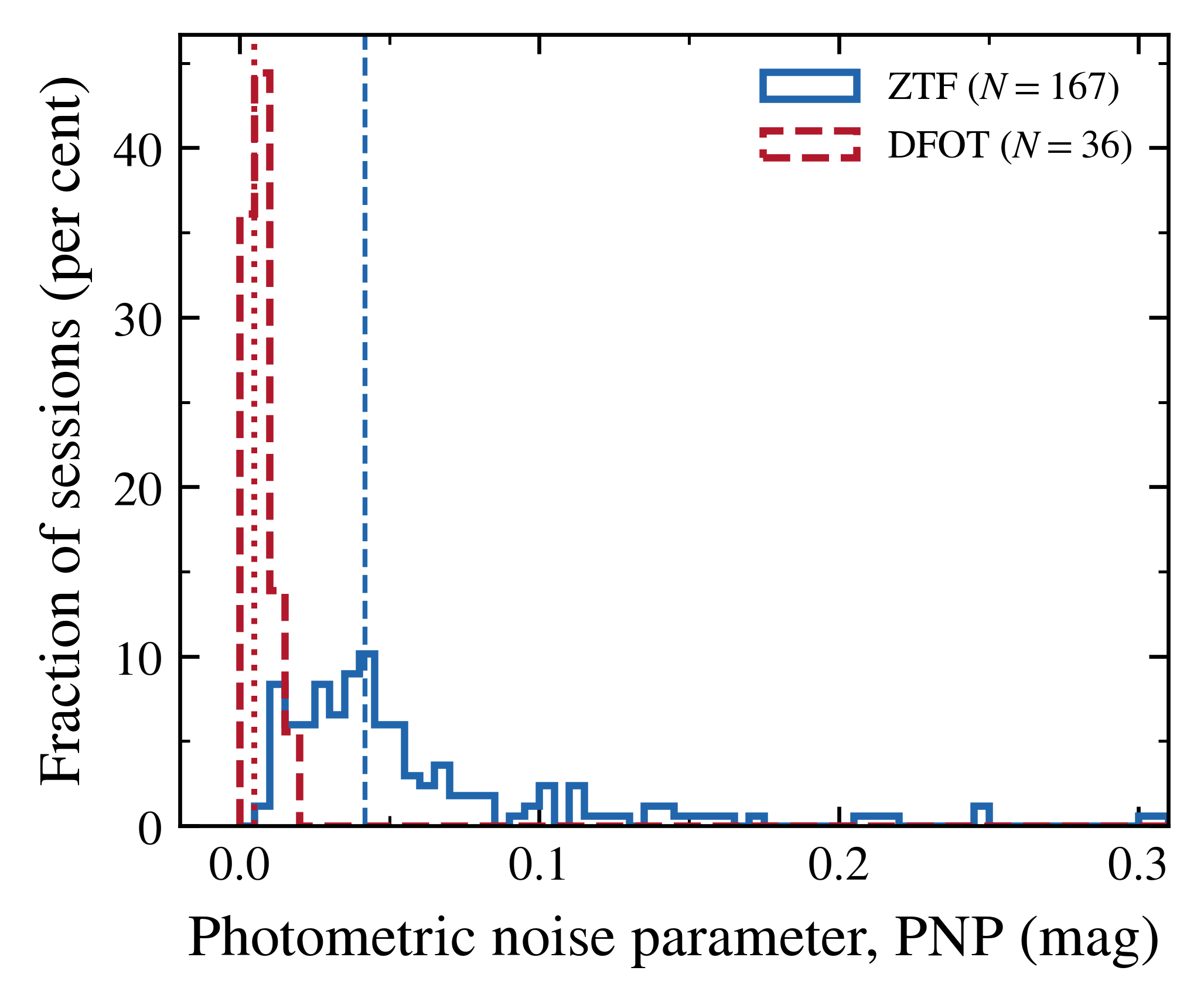}
    \caption{Distributions of the photometric noise parameter (PNP) for the 167 ZTF observing sessions analysed in the present work and the 36 DFOT observing sessions reported in Paper~I. The PNP characterises the typical rms photometric uncertainty of the individual data points in the AGN DLCs after applying the error-scaling factor $\eta=1.54$ (see Section~\ref{subsec:photometric_sensitivity}).}
    \label{fig:pnp_comparison}
\end{figure}

Figure~\ref{fig:pnp_comparison} compares the PNP distributions for the 167 ZTF observing sessions analysed in the present work, with those for the 36 DFOT sessions reported in Paper~I. The contrast between the two distributions is discussed in Section~\ref{sec:discussion}.

\subsection{Statistical assessment of the presence of INOV}
\label{subsec:ftest}

The variability status of the target LMAGN during each session was determined using both its differential light curves (DLCs) using the \(F\)-test \citep{de_Diego_2010}. Following \citet{Goyal2013eta}, we use the error-scaled form of the \(F\)-statistic, denoted by \(F^{\eta}\), which for a given DLC is defined as

\begin{equation}
F^{\eta}=
\frac{{\rm Var}(q)}
{\eta^2 \left(\sum_{i=1}^{N}\sigma_{i,\rm err}^{2}(q)/N\right)},
\end{equation}

where \({\rm Var}(q)\) is the variance of the DLC, \(\sigma_{i,\rm err}(q)\) is the formal photometric uncertainty associated with its \(i\)-th data point, and \(N\) is the number of data points in the accepted (unshaded) segment of the DLC. Here, ($\eta=1.54$) is the same photometric-error scaling factor which was introduced above while defining the PNP. Both the variance and the mean squared photometric uncertainty were computed only for the unshaded portions of the DLCs (see Section~\ref{subsec:quality}). The calculated \textit{$F^{\eta}$} value for a DLC was then compared with the critical values ($F_c$) corresponding to the 99\% confidence level for the appropriate degrees of freedom (see Table~\ref{tab:inov_results} for details). An AGN in an observing session would be classified as Variable (V) when both the Q--S1 and Q--S2 DLCs satisfied $F^{\eta} \geq F_{\rm c}(0.99)$ and, at the same time, the S1--S2 DLC showed no variability. 

\begin{figure*}
    \centering
    \includegraphics[width=\textwidth]{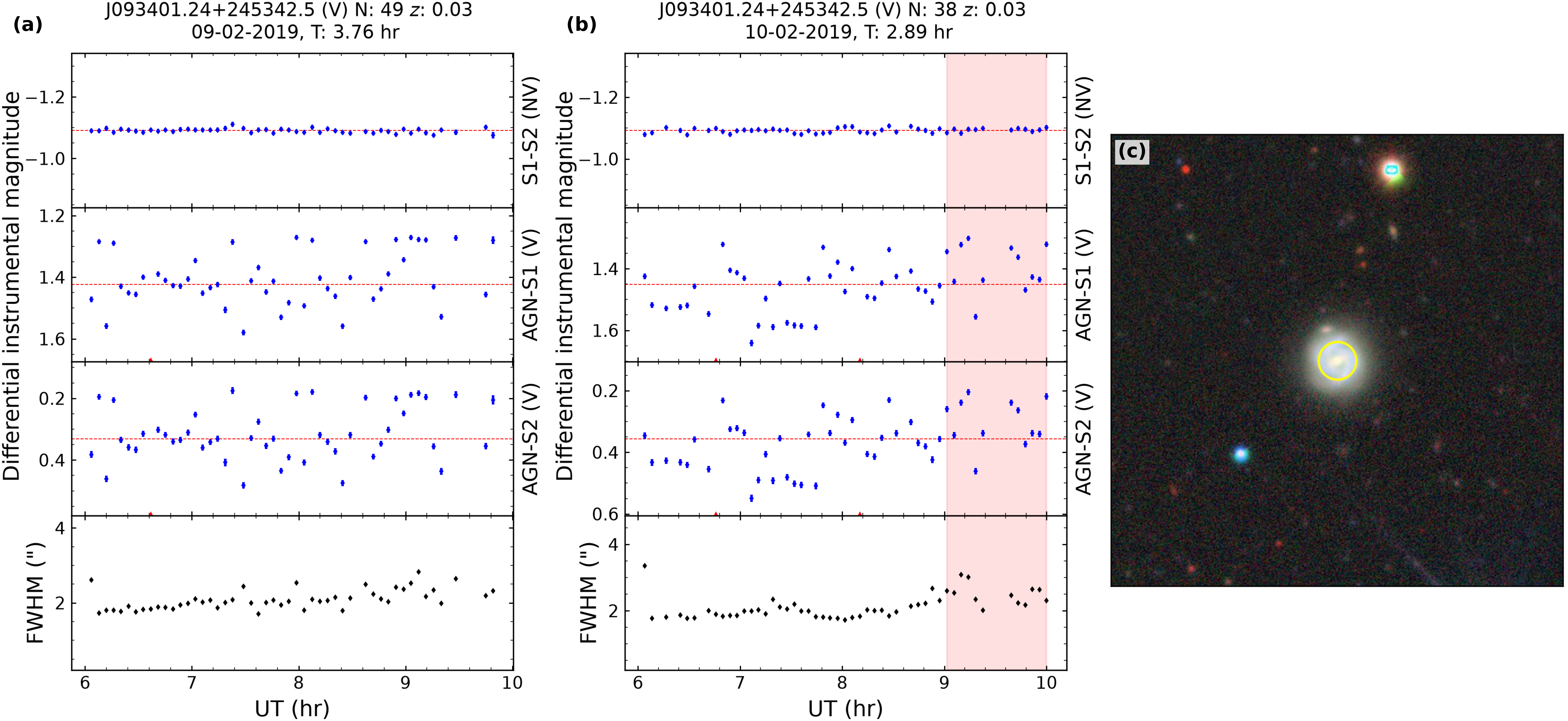}
    \caption{Intranight differential light curves (DLCs) for the two of the intranight monitoring sessions of the LMAGN J093401.24+245342.5 that yielded statistically highly significant (albeit probably spurious) detections of INOV, together with its DECaLS image. Panels (a) \& (b) show, from top to bottom, the `star--star' DLC, the two `LMAGN--star' DLCs, and the variation of the seeing disk (i.e. PSF) during the session. The shaded time segament marks the data excluded from the statistical analysis (Section~\ref{subsec:quality}). Panel (c) displays the DECaLS image cutout measuring $75\,\mathrm{arcsec}$ on each side, centred on J093401.24+245342.5. Although the `LMAGN--star' DLCs show highly significant strong fluctuations under fairly stable PSF conditions, the apparently non-point-like morphology of the central source leads to the strong possibility that these INOV detections with high statistical significance may, in fact, be spurious (see Section~\ref{sec:discussion}).}
    \label{fig:j0934_candidate_inov}
\end{figure*}

Application of the above criterion to the final sample of 167 observing sessions yielded highly significant INOV detection for the first two sessions of J093401.24+245342.5 shown in Fig.~\ref{fig:j0934_candidate_inov}; however, as discussed in Section~\ref{sec:discussion}, the genuineness of these statistically significant detections is doubtful because of the complex morphology of the target LMAGN, as discussed in the next section.

\section{Discussion}
\label{sec:discussion}

The main question addressed in this work, as also in Paper~I, is whether IMBHs, which are supposed to reside at the cores of low-mass galaxies, are capable of ejecting relativistic jets (see, e.g., \citealt{Gopal_witta2008}). An important, widely used diagnostic of such activity is optical variability (see, e.g., \citealt{urry1995}; \citealt{Padovani_2017}; \citealt{blandford2019}). Arguably, intranight optical variability (INOV) is an even more potent diagnostic of jet activity specially, for low-mass AGN (e.g., \citealt{GopalKrishnaWiita2018}) where AGN signatures are often hard to disentangle from starburst activity (Section~\ref{sec:intro}). In Paper~I we presented some evidence for INOV in a sample of IMBHs (LMAGN), although the rather small sample size called for an independent INOV search using a much larger sample of IMBHs.

With the above objective, we have analysed here a well-defined set of 167 observing sessions extracted from the $r$-band archival data from the ZTF survey. These sessions pertain to 62 LMAGNs and each session has an almost continuous sampling over at least 2 hours (median 3.2 hr), considering only those time segments over which the seeing disk i.e., PSF remained fairly steady (Section~\ref{subsec:quality}). The archival ZTF data for each session was subjected to a uniform analysis procedure involving fixed-aperture (for a session) photometry and generation of differential light curves (DLCs) of the target IMBH, relative to two simultaneously monitored comparison stars carefully selected by us for each session, particularly, after verifying their steadiness through the session. The same database was also used to determine the variation of the PSF, for each session (see Fig. \ref{fig:j0934_candidate_inov} for an example of the plots). This cautionary approach, namely, the steadiness of PSF, also adopted in Paper~I, is particularly relevant for LMAGN since in all these low-$z$ targets the emission from the host galaxy is a significant contributor to the aperture photometry of the LMAGN and the degree of such contamination would change in case the PSF varies. Depending on the aperture size used, this could lead to a statistically significant, yet spurious INOV detection (Section~\ref{subsec:quality}).

It is interesting to note that the DLCs derived here from the simultaneously acquired photometric sequences of the target LMAGN and the selected pair of (steady) comparison stars registered in the same sequence of CCD frames, frequently offer a significant, if not dramatic, improvement in the overall quality \textit{vis a vis} the archival flux calibrated ZTF light curves which are based on PSF photometry of the same data. To highlight this important point, we display in Figs.~\ref{fig:j0731_comparison}--\ref{fig:j0830_comparison} a comparison of the two aforementioned sets of light curves, for 4 LMAGN belonging to our sample. In each session, the ZTF-psf light curve is seen to exhibit strong INOV, while no such signature is evident in the corresponding DLC obtained in the present work, neither from visual inspection, nor by applying the $F^{\eta}$-test (see Table~\ref{tab:inov_results}). Importantly, such discrepancies found for these {\it intra-night} light curves are not limited to sessions affected by a conspicuous PSF variation (see, e.g., Fig~\ref{fig:j0731_comparison}). This suggests that PSF-fitting/model systematics associated with the spatially extended host-galaxy emission can also introduce apparent variability into the ZTF-psf photometry of these nearby LMAGNs. We note that very recently, \citealt{Arevalo2026} have also pointed out a similar discrepancy with regard to {\it long-term} light curves of nearby AGN. This discrepancy stood out in their comparison of many ZTF-psf long-term light curves of nearby (\textit{z} < 0.5) AGN taken from the ZTF database, with the corresponding light curves which they generated by subtracting the ZTF `reference image' from the sequence of ZTF science images constituting those light curves.

For only one LMAGN, J093401.24+245342.5, the statistical analysis summarised in Table~\ref{tab:inov_results} suggests a very high significance of INOV detection during the first two sessions. During these sessions, PSF remained fairly stable and clearly, does not reflect the large point-to-point fluctuations present in the DLCs of the LMAGN (Figure~\ref{fig:j0934_candidate_inov}). However, a closer scrutiny of the CCD frames raises doubts about the reality of these INOV signatures. This is because the operation of the aperture-photometry algorithm DAOPHOT appears to be severly compromised in this case, due to the confusion caused by additional emission peak(s) present near the nucleus. Indeed, a close inspection of the DECaLS image in Figure~\ref{fig:j0934_candidate_inov} indicates deviation from a point-like central peak. 

The absence of INOV detection in the present large sample of LMAGN will now be discussed in the context of the rather high INOV detection rate reported for LMAGN in Paper~I. As a first step, we shall make a the statistical comparison of the rms noise for the present set of 167 ZTF-based intranight sessions with the corresponding noise statistics for the 36 DFOT-based intranight sessions reported in Paper~I. The basic parameter we are comparing is the session PNP, defined in Section~\ref{subsec:photometric_sensitivity}. The two histograms in Figure~\ref{fig:pnp_comparison} show that the PNP values for all the sessions pertaining to Paper~I are $<0.02$ mag (median = 0.005 mag), whereas the corresponding distribution for the present ZTF-based sample, which was observed in a survey mode, is much broader, peaking near 0.045 mag and having a long tail extending up to $\sim0.3$ mag (median $\sim0.041$ mag). In fact, there seems to be only a slight overlap between the two PNP distributions. Thus, even in terms of median value, the DFOT-based sample, which was monitored in a targeted mode (Paper~I) had $\sim$ 8 times higher sensitivity. Part of this sensitivity contrast may have arisen from the difference between the typical exposure time of individual data points, being 0.5 minutes for the ZTF and $\sim5$ minutes for the DFOT data points. Accounting approximately for the expected dependence of the photometric noise on exposure time still leaves a factor of $\sim2.5$ (statistically) by which the present ZTF-based large sample of DLCs is less sensitive than DFOT-based sample presented in Paper~I. The contrast between the INOV duty cycles—nearly zero for the present IMBH sample, compared to INOV  duty cycle of nearly $1/5$ as reported in Paper~I may therefore be attributed partly to this sensitivity factor.
\begin{figure}
    \centering
    \includegraphics[width=\columnwidth]{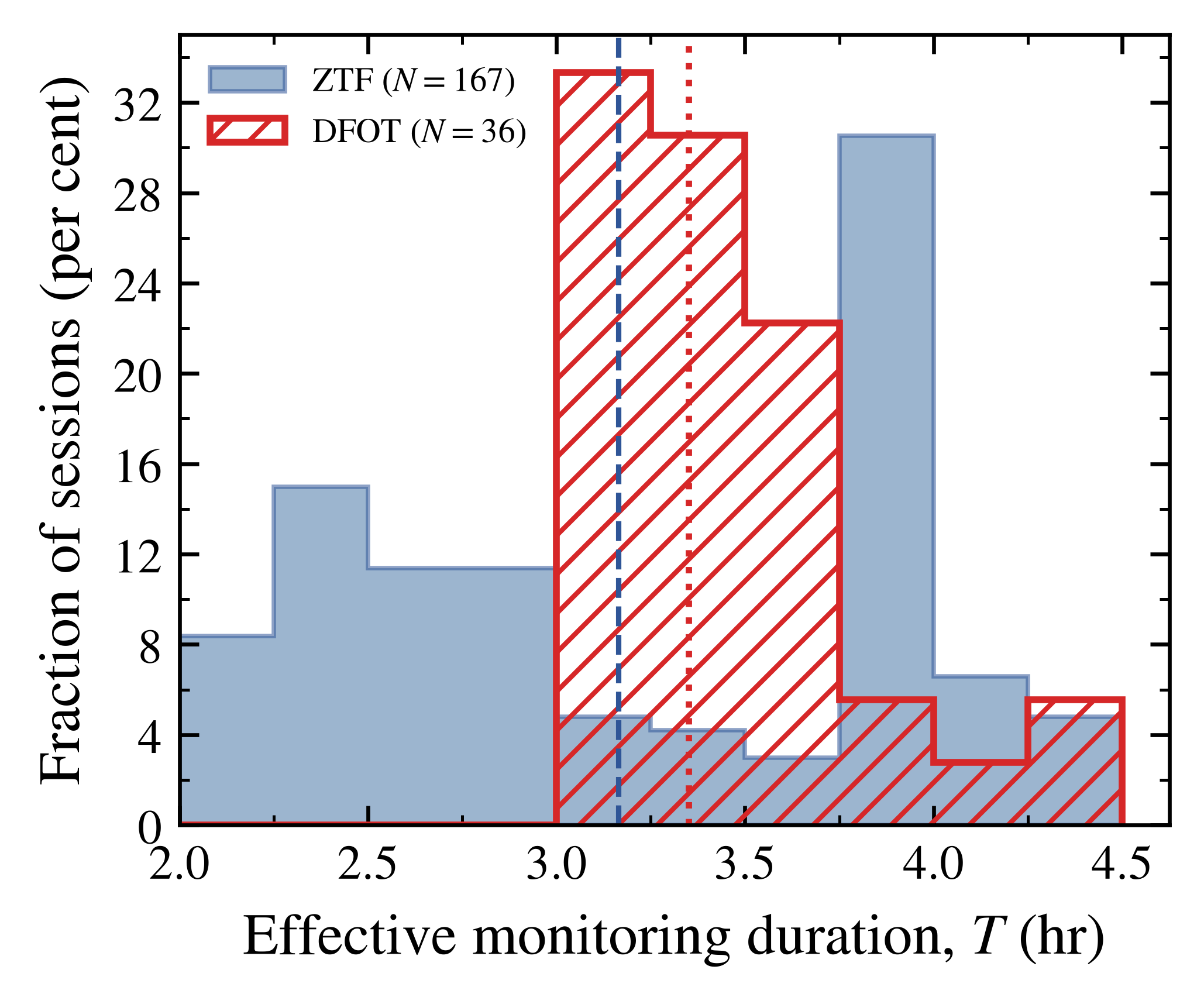}
    \caption{
    Distributions of the effective monitoring duration, $T$, for the 167 ZTF-based intranight sessions analysed in the present study and the 36 DFOT-based sessions reported in Paper~I. Each histogram is normalised to the total number of sessions in the respective sample, and both distributions are plotted using a common bin width of 0.25\,hr. The blue dashed and red dotted vertical lines mark the median durations of the ZTF and DFOT samples, respectively.
    }
    \label{fig:duration_hist}
\end{figure}
The duration of an observing session (T) is another factor governing the probability of detecting INOV, which is known to slowly increase with T (e.g. \citealt{Carini2007} and references therein). Figure~\ref{fig:duration_hist} compares the distributions of the effective monitoring duration for the present ZTF-based sample and the DFOT-based sample reported in Paper~I. Median values of T for the two samples are quite similar, being 3.20\,hr and 3.35\,hr for the two samples, respectively. The corresponding mean durations are also nearly the same (3.24\,hr and 3.44\,hr). On the other hand, the two histograms look strikingly different in shape. Whereas T lies between 3.0\,hr and 3.75\,hr for bulk of the DFOT sample, almost half of the ZTF sample has T values below this range and only a small minority falls on the higher side of this range (Figure~\ref{fig:duration_hist}). This too might have contributed to the lack of INOV detection found here for the ZTF sample, albeit to a lesser degree compared to the sensitivity factor mentioned above. We note that the telescope sizes used for the two studies are very similar (1.2~m for the ZTF survey telescope and 1.3~m for the DFOT).

Some additional differences include the use in Paper~I of an IMBH sample which differs from the present sample, not just in being several times smaller (hence yielding a relatively weaker statistics) but also in terms of the sample selection criterion adopted in Paper~I which mandated a confirmed detection of each selected IMBH in both radio and X-ray bands. Of course, the possibility of some unrecognised observational issues having also played some role cannot be discounted.

\section{Conclusions}
\label{sec:conclusions}

In this work we have attempted to search for intranight optical variability among AGN which are believed to be powered by intermediate-mass black holes (IMBH) having masses ($M_{BH}$) in the range $8\times10^{4}$ to $ 2\times10^{6}\,M_\odot$. To build a large sample of such low-mass AGN (LMAGN), we subjeted their published two large catalogues to a selection process aimed at enhancing suitability for INOV search, while avoiding to introducing any biases. The application of well-motivated selection filters to the  LMAGN catalogues and to their $r$-band intranight science images available in the ZTF database, has led to a final sample containing 62 LMAGN in the redshift range 0.01 < $z$ < 0.30, which had been monitored in 167 ZTF sessions of duration > 2 hours and marked by a fairly steady seeing disk (PSF). In a fresh analysis, we subjected these data to fixed aperture photometry, also paying attention to steadiness of the seeing disk during individual sessions. By comparing our photometric measurements of the LMAGNs with those made by us for two carefully chosen comparison stars that were monitored simultaneously to the target  LMAGN in the same sequence of ZTF science images, two differential light curves (DLCs) of the LMAGN were derived for each session. Application of the widely used statistical test based on $F$-statistics to these DLCs has revealed no case of robust INOV detection. We have discussed this finding in the context of an earlier report of INOV detections in a much smaller sample of IMBH, which was, however, assembled using different selection criteria and monitored in a significantly more sensitive, targeted campaign. 

\begin{table*}

    \centering

    \caption{Results of the INOV analysis for the 167 ZTF intranight sessions. The subscripts 1 and 2 correspond to the Q--S1 and Q--S2 differential light curves.}
    \label{tab:inov_results}

    \scriptsize

    \setlength{\tabcolsep}{2pt}
    \renewcommand{\arraystretch}{1.05}

    \resizebox{\textwidth}{!}{%
    \begin{tabular}{llcccccc@{\hspace{5mm}}llcccccc}

    \cmidrule(lr){1-8}
    \cmidrule(lr){9-16}

    Source & Date & $z$ & $N$ & $T$ & $F_c^{0.99}$ & $F^{\eta}_1$ & $F^{\eta}_2$ & Source & Date & $z$ & $N$ & $T$ & $F_c^{0.99}$ & $F^{\eta}_1$ & $F^{\eta}_2$ \\

    \cmidrule(lr){1-8}
    \cmidrule(lr){9-16}

    J073106.87+392644.7 & 01-01-2020 & 0.049 & 55 & 3.93 & 1.89 & 0.30 & 0.27 & J080537.37+362522.1 & 01-01-2020 & 0.088 & 51 & 3.95 & 1.94 & 0.44 & 0.42 \\
J073106.87+392644.7 & 03-01-2020 & 0.049 & 50 & 4.29 & 1.95 & 0.51 & 0.37 & J080537.37+362522.1 & 03-01-2020 & 0.088 & 48 & 3.94 & 1.98 & 0.33 & 0.33 \\
J073106.87+392644.7 & 05-01-2020 & 0.049 & 57 & 4.20 & 1.87 & 0.25 & 0.23 & J080537.37+362522.1 & 05-01-2020 & 0.088 & 22 & 2.26 & 2.78 & 0.26 & 0.26 \\
J073106.87+392644.7 & 14-11-2019 & 0.049 & 47 & 2.18 & 1.99 & 0.22 & 0.22 & J080537.37+362522.1 & 20-12-2019 & 0.088 & 55 & 4.35 & 1.89 & 0.50 & 0.47 \\
J073106.87+392644.7 & 20-12-2019 & 0.049 & 55 & 3.89 & 1.89 & 0.29 & 0.48 & J080537.37+362522.1 & 28-01-2019 & 0.088 & 30 & 2.91 & 2.39 & 0.23 & 0.30 \\
J073106.87+392644.7 & 28-01-2019 & 0.049 & 32 & 2.91 & 2.32 & 0.21 & 0.16 & J080537.37+362522.1 & 29-12-2019 & 0.088 & 37 & 2.80 & 2.18 & 0.50 & 0.54 \\
J073106.87+392644.7 & 29-12-2019 & 0.049 & 39 & 2.79 & 2.14 & 0.38 & 0.29 & J080743.73+345454.4 & 01-01-2020 & 0.108 & 52 & 3.95 & 1.92 & 0.34 & 0.44 \\
J074251.09+333403.9 & 01-01-2020 & 0.126 & 48 & 3.83 & 1.98 & 0.18 & 0.21 & J080743.73+345454.4 & 03-01-2020 & 0.108 & 48 & 4.09 & 1.98 & 0.64 & 0.62 \\
J074251.09+333403.9 & 03-01-2020 & 0.126 & 40 & 4.25 & 2.11 & 0.32 & 0.29 & J080743.73+345454.4 & 05-01-2020 & 0.108 & 26 & 2.44 & 2.55 & 0.22 & 0.23 \\
J074251.09+333403.9 & 05-01-2020 & 0.126 & 49 & 4.10 & 1.96 & 0.34 & 0.25 & J080743.73+345454.4 & 20-12-2019 & 0.108 & 55 & 4.35 & 1.89 & 0.23 & 0.34 \\
J074251.09+333403.9 & 20-12-2019 & 0.126 & 48 & 3.88 & 1.98 & 0.35 & 0.47 & J080743.73+345454.4 & 28-01-2019 & 0.108 & 31 & 2.91 & 2.35 & 0.27 & 0.21 \\
J074251.09+333403.9 & 28-01-2019 & 0.126 & 39 & 3.91 & 2.14 & 0.34 & 0.38 & J080743.73+345454.4 & 29-12-2019 & 0.108 & 37 & 2.80 & 2.18 & 0.39 & 0.34 \\
J074251.09+333403.9 & 29-12-2019 & 0.126 & 35 & 2.80 & 2.23 & 0.23 & 0.25 & J080910.72+110619.1 & 04-01-2020 & 0.053 & 45 & 2.74 & 2.02 & 0.21 & 0.29 \\
J074423.45+243046.3 & 03-01-2020 & 0.117 & 52 & 4.25 & 1.92 & 0.39 & 0.38 & J080910.72+110619.1 & 10-12-2019 & 0.053 & 29 & 2.32 & 2.42 & 0.29 & 0.23 \\
J074423.45+243046.3 & 10-11-2019 & 0.117 & 57 & 2.21 & 1.87 & 0.27 & 0.34 & J080910.72+110619.1 & 15-11-2019 & 0.053 & 61 & 2.01 & 1.83 & 0.25 & 0.28 \\
J074423.45+243046.3 & 20-12-2019 & 0.117 & 52 & 3.93 & 1.92 & 0.46 & 0.66 & J080910.72+110619.1 & 28-01-2019 & 0.053 & 28 & 2.47 & 2.46 & 0.30 & 0.26 \\
J074423.45+243046.3 & 29-12-2019 & 0.117 & 29 & 2.38 & 2.42 & 0.57 & 0.54 & J081550.24+250641.0 & 01-01-2020 & 0.073 & 54 & 3.86 & 1.90 & 0.25 & 0.24 \\
J074659.82+440728.8 & 01-01-2020 & 0.076 & 33 & 2.38 & 2.29 & 0.34 & 0.28 & J081550.24+250641.0 & 03-01-2020 & 0.073 & 45 & 3.95 & 2.02 & 0.44 & 0.29 \\
J074659.82+440728.8 & 03-01-2020 & 0.076 & 49 & 4.11 & 1.96 & 0.72 & 0.67 & J081550.24+250641.0 & 28-01-2019 & 0.073 & 31 & 2.91 & 2.35 & 0.34 & 0.40 \\
J074659.82+440728.8 & 05-01-2020 & 0.076 & 28 & 2.24 & 2.46 & 0.37 & 0.22 & J081933.38+231704.7 & 01-01-2020 & 0.094 & 52 & 3.86 & 1.92 & 0.21 & 0.19 \\
J074659.82+440728.8 & 14-11-2019 & 0.076 & 46 & 2.18 & 2.01 & 0.61 & 0.61 & J081933.38+231704.7 & 03-01-2020 & 0.094 & 44 & 4.20 & 2.04 & 1.18 & 1.28 \\
J074659.82+440728.8 & 20-12-2019 & 0.076 & 57 & 4.35 & 1.87 & 0.56 & 0.65 & J081933.38+231704.7 & 05-01-2020 & 0.094 & 38 & 3.86 & 2.16 & 1.15 & 1.23 \\
J074659.82+440728.8 & 28-01-2019 & 0.076 & 39 & 3.86 & 2.14 & 0.83 & 0.97 & J081933.38+231704.7 & 28-01-2019 & 0.094 & 25 & 2.49 & 2.60 & 0.28 & 0.46 \\
J074659.82+440728.8 & 29-12-2019 & 0.076 & 39 & 2.79 & 2.14 & 0.85 & 0.89 & J082325.92+065106.5 & 04-01-2020 & 0.072 & 46 & 2.74 & 2.01 & 0.33 & 0.28 \\
J074948.33+264734.2 & 03-01-2020 & 0.132 & 50 & 4.01 & 1.95 & 0.53 & 0.82 & J082325.92+065106.5 & 07-11-2019 & 0.072 & 58 & 2.18 & 1.86 & 0.18 & 0.31 \\
J074948.33+264734.2 & 10-11-2019 & 0.132 & 50 & 2.21 & 1.95 & 0.19 & 0.23 & J082502.16+082614.9 & 16-11-2019 & 0.082 & 47 & 2.07 & 1.99 & 1.09 & 0.97 \\
J074948.33+264734.2 & 20-12-2019 & 0.132 & 54 & 3.93 & 1.90 & 0.36 & 0.40 & J082502.99+111021.4 & 16-11-2019 & 0.108 & 48 & 2.10 & 1.98 & 1.57 & 1.43 \\
J074948.33+264734.2 & 29-12-2019 & 0.132 & 30 & 2.38 & 2.39 & 0.42 & 0.33 & J082741.35+175751.0 & 04-01-2020 & 0.107 & 34 & 2.34 & 2.26 & 0.19 & 0.17 \\
J075709.34+423616.4 & 01-01-2020 & 0.074 & 33 & 2.38 & 2.29 & 0.46 & 0.26 & J082741.35+175751.0 & 28-01-2019 & 0.107 & 30 & 2.65 & 2.39 & 0.34 & 0.27 \\
J075709.34+423616.4 & 03-01-2020 & 0.074 & 47 & 3.99 & 1.99 & 0.55 & 0.36 & J083021.81+183031.2 & 04-01-2020 & 0.097 & 41 & 2.62 & 2.09 & 0.27 & 0.27 \\
J075709.34+423616.4 & 05-01-2020 & 0.074 & 26 & 2.24 & 2.55 & 0.36 & 0.23 & J083021.81+183031.2 & 08-11-2019 & 0.097 & 46 & 2.06 & 2.01 & 0.22 & 0.14 \\
J075709.34+423616.4 & 14-11-2019 & 0.074 & 69 & 2.03 & 1.76 & 0.28 & 0.25 & J083021.81+183031.2 & 11-12-2019 & 0.097 & 28 & 2.35 & 2.46 & 0.62 & 0.69 \\
J075709.34+423616.4 & 20-12-2019 & 0.074 & 57 & 4.35 & 1.87 & 0.64 & 0.58 & J083021.81+183031.2 & 28-01-2019 & 0.097 & 30 & 2.65 & 2.39 & 0.27 & 0.19 \\
J075709.34+423616.4 & 28-01-2019 & 0.074 & 42 & 3.91 & 2.08 & 0.66 & 0.52 & J083916.92+174327.5 & 09-11-2019 & 0.105 & 50 & 2.20 & 1.95 & 0.31 & 0.41 \\
J075709.34+423616.4 & 29-12-2019 & 0.074 & 39 & 2.79 & 2.14 & 0.50 & 0.57 & J084303.60+361011.2 & 08-02-2019 & 0.087 & 39 & 4.45 & 2.14 & 0.35 & 0.29 \\
J080352.99+263123.4 & 01-01-2020 & 0.046 & 52 & 3.86 & 1.92 & 0.41 & 0.45 & J084739.61+360444.4 & 08-02-2019 & 0.087 & 54 & 3.99 & 1.90 & 0.29 & 0.23 \\
J080352.99+263123.4 & 03-01-2020 & 0.046 & 47 & 4.22 & 1.99 & 0.98 & 0.81 & J084948.56+150636.7 & 08-02-2019 & 0.071 & 54 & 3.92 & 1.90 & 0.29 & 0.20 \\
J080352.99+263123.4 & 20-12-2019 & 0.046 & 49 & 3.95 & 1.96 & 1.54 & 1.14 & J084948.56+150636.7 & 13-12-2019 & 0.071 & 36 & 2.58 & 2.21 & 0.37 & 0.44 \\
J080352.99+263123.4 & 29-12-2019 & 0.046 & 37 & 2.57 & 2.18 & 0.59 & 0.51 & J084948.56+150636.7 & 16-11-2019 & 0.071 & 51 & 2.26 & 1.94 & 0.31 & 0.29 \\
J080511.91+064119.5 & 02-01-2020 & 0.100 & 36 & 3.43 & 2.21 & 0.35 & 0.32 & J085152.63+522833.0 & 24-01-2019 & 0.065 & 56 & 3.94 & 1.88 & 0.71 & 0.69 \\
J080511.91+064119.5 & 04-01-2020 & 0.100 & 50 & 2.90 & 1.95 & 0.22 & 0.23 & J085152.63+522833.0 & 26-01-2019 & 0.065 & 55 & 3.95 & 1.89 & 0.45 & 0.42 \\
J080511.91+064119.5 & 07-11-2019 & 0.100 & 63 & 2.18 & 1.81 & 0.30 & 0.34 & J085655.29+442653.4 & 24-01-2019 & 0.179 & 45 & 3.44 & 2.02 & 0.34 & 0.38 \\

    \cmidrule(lr){1-8}
    \cmidrule(lr){9-16}

    \end{tabular}%
    }

    \end{table*}

\begin{table*}

    \ContinuedFloat

    \centering

    \caption{continued}

    \scriptsize

    \setlength{\tabcolsep}{2pt}
    \renewcommand{\arraystretch}{1.05}

    \resizebox{\textwidth}{!}{%
    \begin{tabular}{llcccccc@{\hspace{5mm}}llcccccc}

    \cmidrule(lr){1-8}
    \cmidrule(lr){9-16}

    Source & Date & $z$ & $N$ & $T$ & $F_c^{0.99}$ & $F^{\eta}_1$ & $F^{\eta}_2$ & Source & Date & $z$ & $N$ & $T$ & $F_c^{0.99}$ & $F^{\eta}_1$ & $F^{\eta}_2$ \\

    \cmidrule(lr){1-8}
    \cmidrule(lr){9-16}

    J085655.29+442653.4 & 26-01-2019 & 0.179 & 40 & 3.88 & 2.11 & 0.34 & 0.31 & J092700.53+084329.5 & 10-02-2019 & 0.112 & 36 & 2.46 & 2.21 & 0.23 & 0.30 \\
J085836.04+353634.3 & 08-02-2019 & 0.126 & 55 & 3.99 & 1.89 & 0.26 & 0.20 & J092700.53+084329.5 & 12-02-2019 & 0.112 & 44 & 3.02 & 2.04 & 0.37 & 0.48 \\
J085906.36+542150.2 & 24-01-2019 & 0.182 & 39 & 3.23 & 2.14 & 0.44 & 0.39 & J092740.77+262701.1 & 09-02-2019 & 0.049 & 42 & 2.92 & 2.08 & 0.35 & 0.47 \\
J085906.36+542150.2 & 26-01-2019 & 0.182 & 30 & 3.88 & 2.39 & 0.26 & 0.34 & J092740.77+262701.1 & 10-02-2019 & 0.049 & 38 & 2.60 & 2.16 & 0.65 & 0.55 \\
J090032.80+255650.8 & 08-02-2019 & 0.051 & 54 & 3.92 & 1.90 & 0.24 & 0.21 & J092740.77+262701.1 & 12-02-2019 & 0.049 & 43 & 2.95 & 2.06 & 0.50 & 0.40 \\
J090320.97+045738.1 & 08-02-2019 & 0.057 & 52 & 3.92 & 1.92 & 0.19 & 0.20 & J093308.89+534749.0 & 24-01-2019 & 0.057 & 53 & 3.94 & 1.91 & 0.33 & 0.24 \\
J090320.97+045738.1 & 10-11-2019 & 0.057 & 50 & 2.28 & 1.95 & 0.28 & 0.20 & J093308.89+534749.0 & 26-01-2019 & 0.057 & 36 & 2.49 & 2.21 & 0.29 & 0.47 \\
J090355.74+283900.7 & 08-02-2019 & 0.148 & 47 & 3.92 & 1.99 & 0.33 & 0.47 & J093401.24+245342.5 & 09-02-2019 & 0.033 & 49 & 3.76 & 1.96 & 56.30 & 39.29 \\
J090431.22+075330.9 & 08-02-2019 & 0.084 & 54 & 3.92 & 1.90 & 0.39 & 0.40 & J093401.24+245342.5 & 10-02-2019 & 0.033 & 38 & 2.89 & 2.16 & 48.44 & 37.09 \\
J090431.22+075330.9 & 10-11-2019 & 0.084 & 51 & 2.28 & 1.94 & 0.43 & 0.33 & J093401.24+245342.5 & 12-02-2019 & 0.033 & 44 & 3.38 & 2.04 & 3.68 & 5.59 \\
J090731.06+445428.5 & 24-01-2019 & 0.116 & 33 & 3.30 & 2.29 & 0.48 & 0.36 & J093408.60+175644.0 & 09-02-2019 & 0.027 & 41 & 2.92 & 2.09 & 0.16 & 0.22 \\
J090731.06+445428.5 & 26-01-2019 & 0.116 & 38 & 3.81 & 2.16 & 0.38 & 0.38 & J093408.60+175644.0 & 10-02-2019 & 0.027 & 38 & 2.60 & 2.16 & 0.50 & 0.40 \\
J091023.11+302729.2 & 09-02-2019 & 0.048 & 40 & 2.92 & 2.11 & 0.21 & 0.27 & J093408.60+175644.0 & 12-02-2019 & 0.027 & 57 & 3.93 & 1.87 & 0.67 & 0.52 \\
J091023.11+302729.2 & 10-02-2019 & 0.048 & 34 & 2.39 & 2.26 & 0.48 & 0.33 & J093829.39+034826.7 & 10-02-2019 & 0.119 & 37 & 2.53 & 2.18 & 0.22 & 0.12 \\
J091023.11+302729.2 & 12-02-2019 & 0.048 & 55 & 3.86 & 1.89 & 0.33 & 0.33 & J093829.39+034826.7 & 12-02-2019 & 0.119 & 45 & 3.09 & 2.02 & 0.43 & 0.48 \\
J091045.91+480339.4 & 24-01-2019 & 0.106 & 54 & 3.94 & 1.90 & 0.44 & 0.27 & J093946.86+143606.7 & 10-02-2019 & 0.190 & 39 & 2.67 & 2.14 & 0.41 & 0.35 \\
J091045.91+480339.4 & 26-01-2019 & 0.106 & 54 & 3.88 & 1.90 & 0.38 & 0.40 & J093946.86+143606.7 & 12-02-2019 & 0.190 & 31 & 2.74 & 2.35 & 0.18 & 0.18 \\
J091424.76+115625.6 & 08-02-2019 & 0.031 & 38 & 3.85 & 2.16 & 0.33 & 0.35 & J094057.20+032401.2 & 10-02-2019 & 0.061 & 37 & 2.53 & 2.18 & 0.35 & 0.23 \\
J091424.76+115625.6 & 13-12-2019 & 0.031 & 32 & 2.58 & 2.32 & 0.31 & 0.31 & J094057.20+032401.2 & 12-02-2019 & 0.061 & 45 & 3.09 & 2.02 & 0.29 & 0.42 \\
J091449.06+085321.1 & 13-12-2019 & 0.140 & 36 & 2.58 & 2.21 & 0.73 & 0.69 & J094258.55+083810.1 & 10-02-2019 & 0.134 & 38 & 2.60 & 2.16 & 0.28 & 0.33 \\
J091449.06+085321.1 & 16-11-2019 & 0.140 & 51 & 2.26 & 1.94 & 0.63 & 0.66 & J094258.55+083810.1 & 12-02-2019 & 0.134 & 22 & 3.02 & 2.78 & 0.30 & 0.32 \\
J091738.08+333939.0 & 09-02-2019 & 0.173 & 39 & 3.48 & 2.14 & 0.28 & 0.22 & J094529.37+093610.4 & 10-02-2019 & 0.013 & 39 & 2.67 & 2.14 & 0.51 & 0.40 \\
J091738.08+333939.0 & 10-02-2019 & 0.173 & 33 & 3.45 & 2.29 & 0.34 & 0.24 & J094529.37+093610.4 & 12-02-2019 & 0.013 & 55 & 3.93 & 1.89 & 0.54 & 0.78 \\
J091738.08+333939.0 & 12-02-2019 & 0.173 & 18 & 3.16 & 3.13 & 0.49 & 0.49 & J095151.82+060143.7 & 10-01-2024 & 0.093 & 64 & 2.50 & 1.80 & 0.31 & 0.36 \\
J091858.97+053844.3 & 08-02-2019 & 0.191 & 50 & 3.63 & 1.95 & 0.21 & 0.26 & J095330.54+562653.5 & 24-01-2019 & 0.066 & 56 & 3.94 & 1.88 & 0.40 & 0.42 \\
J091858.97+053844.3 & 10-11-2019 & 0.191 & 65 & 2.28 & 1.79 & 0.31 & 0.29 & J095330.54+562653.5 & 26-01-2019 & 0.066 & 40 & 2.77 & 2.11 & 0.38 & 0.40 \\
J091917.89+050953.0 & 08-02-2019 & 0.075 & 52 & 3.92 & 1.92 & 0.58 & 0.44 & J095553.98+534846.1 & 24-01-2019 & 0.304 & 32 & 3.51 & 2.32 & 0.27 & 0.26 \\
J091917.89+050953.0 & 10-11-2019 & 0.075 & 64 & 2.28 & 1.80 & 0.43 & 0.33 & J095553.98+534846.1 & 26-01-2019 & 0.304 & 43 & 4.20 & 2.06 & 0.31 & 0.28 \\
J092027.85+304858.2 & 09-02-2019 & 0.101 & 46 & 3.83 & 2.01 & 0.54 & 0.58 & J100510.52+543255.5 & 24-01-2019 & 0.050 & 56 & 3.94 & 1.88 & 0.52 & 0.49 \\
J092027.85+304858.2 & 10-02-2019 & 0.101 & 42 & 2.89 & 2.08 & 8.97 & 7.90 & J100510.52+543255.5 & 26-01-2019 & 0.050 & 36 & 2.49 & 2.21 & 0.56 & 0.64 \\
J092027.85+304858.2 & 12-02-2019 & 0.101 & 40 & 3.16 & 2.11 & 0.30 & 0.28 & J100711.44+581714.3 & 24-01-2019 & 0.202 & 33 & 3.75 & 2.29 & 0.39 & 0.42 \\
J092115.45+254911.5 & 09-02-2019 & 0.249 & 50 & 3.69 & 1.95 & 0.30 & 0.31 & J100711.44+581714.3 & 26-01-2019 & 0.202 & 36 & 3.88 & 2.21 & 0.17 & 0.13 \\
J092115.45+254911.5 & 10-02-2019 & 0.249 & 54 & 3.93 & 1.90 & 0.36 & 0.32 & J102703.91+485023.9 & 24-01-2019 & 0.147 & 55 & 3.87 & 1.89 & 0.52 & 0.41 \\
J092115.45+254911.5 & 12-02-2019 & 0.249 & 46 & 3.86 & 2.01 & 0.32 & 0.36 & J102703.91+485023.9 & 25-01-2019 & 0.147 & 55 & 4.13 & 1.89 & 0.50 & 0.44 \\
J092355.85+282637.5 & 09-02-2019 & 0.082 & 51 & 3.69 & 1.94 & 0.28 & 0.36 & J102703.91+485023.9 & 26-01-2019 & 0.147 & 34 & 2.42 & 2.26 & 1.63 & 1.73 \\
J092355.85+282637.5 & 10-02-2019 & 0.082 & 38 & 2.60 & 2.16 & 0.37 & 0.23 & J103928.73+562058.2 & 24-01-2019 & 0.075 & 53 & 3.94 & 1.91 & 0.35 & 0.36 \\
J092355.85+282637.5 & 12-02-2019 & 0.082 & 43 & 2.95 & 2.06 & 0.40 & 0.27 & J103928.73+562058.2 & 25-01-2019 & 0.075 & 56 & 4.25 & 1.88 & 0.32 & 0.28 \\
J092411.46+022321.8 & 08-02-2019 & 0.089 & 50 & 3.69 & 1.95 & 0.28 & 0.23 & J103928.73+562058.2 & 26-01-2019 & 0.075 & 34 & 2.50 & 2.26 & 0.54 & 0.51 \\
J092522.39+130538.7 & 10-02-2019 & 0.080 & 38 & 2.60 & 2.16 & 0.70 & 0.56 & J105131.91+504223.2 & 24-01-2019 & 0.132 & 52 & 3.87 & 1.92 & 0.36 & 0.42 \\
J092522.39+130538.7 & 12-02-2019 & 0.080 & 44 & 3.02 & 2.04 & 0.51 & 0.34 & J105131.91+504223.2 & 25-01-2019 & 0.132 & 56 & 4.47 & 1.88 & 0.29 & 0.27 \\
J092547.32+050231.7 & 10-02-2019 & 0.127 & 35 & 2.39 & 2.23 & 0.52 & 0.48 & J105131.91+504223.2 & 26-01-2019 & 0.132 & 44 & 3.88 & 2.04 & 0.62 & 0.71 \\
J092547.32+050231.7 & 12-02-2019 & 0.127 & 46 & 3.30 & 2.01 & 0.35 & 0.35 &  &  &  &  &  &  &  &  \\

    \cmidrule(lr){1-8}
    \cmidrule(lr){9-16}

    \end{tabular}%
    }

    \end{table*}

% \input{table2_INOV_results_part1}
% \input{table2_INOV_results_part2}
% \input{table2_INOV_results_part3}
% \input{table2_INOV_results_part4}
% \input{table2_INOV_results_part5}
%%%%%%%%%%%%%%%%%%%%%%%%%%%%%%%%%%%%%%%%%%%%%%%%%%
% \clearpage
\section*{Acknowledgements}

HS, HC and AKS are grateful to the Inter-University Centre for Astronomy and Astrophysics (IUCAA) for their hospitality and the provision of High-Performance Computing (HPC) facilities under the IUCAA Associate Programme. GK would like to thank Indian National Science Academy for the award of Senior Scientist position during which part of this work was carried out. VN is supported by Beijing Natural Science Foundation (Grant No. IS25004). KC acknowledges the support of Anusandhan National Research Foundation (ANRF), formerly Science and Engineering Research Board (SERB)-Department of Science and Technology (DST), New Delhi, for funding under the National Post-Doctoral Fellowship Scheme through grant no. PDF/2023/004071.

%%%%%%%%%%%%%%%%%%%%%%%%%%%%%%%%%%%%%%%%%%%%%%%%%%
\section*{Data Availability}

The archival data used in this study are publicly available from the Zwicky Transient Facility (ZTF; \url{https://www.ztf.caltech.edu/}).

%%%%%%%%%%%%%%%%%%%%%%%%%%%%%%%%%%%%%%%%%%%%%%%%%%

% \FloatBarrier
\bibliographystyle{mnras}
\bibliography{example}
\appendix
\onecolumn
\section{Comparison of ZTF PSF-based and aperture photometry}
\label{app:photometry_comparison}
\clearpage
% J0731
\begin{figure}
    \centering
    \includegraphics[width=0.95\textwidth, height=0.72\textheight,
        keepaspectratio]{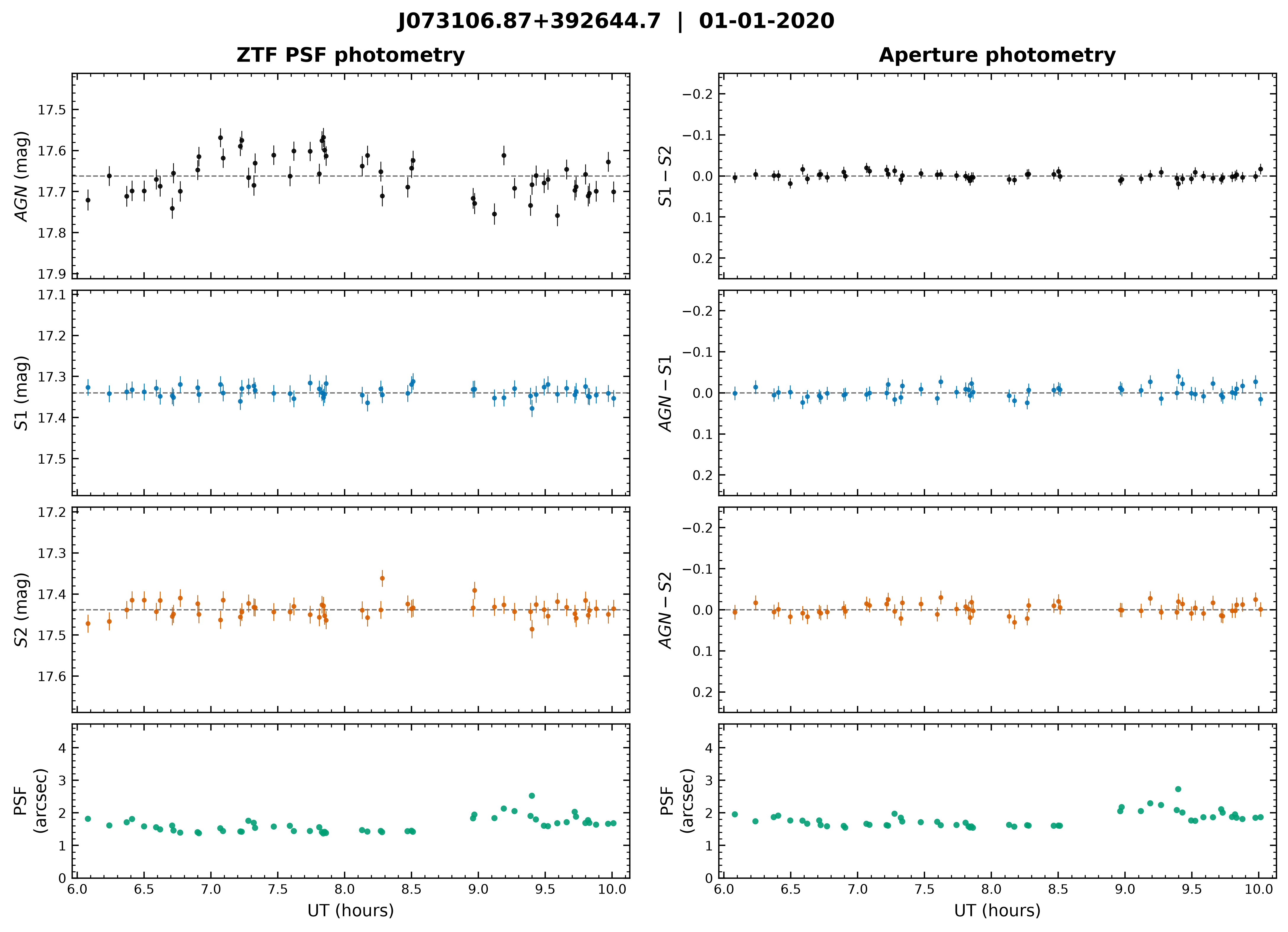}
    \caption{
    Comparison of the ZTF archival PSF-based photometry (left panels) and our independently derived aperture photometry (right panels) for the LMAGN J073106.87+392644.7, observed on 1 January 2020. From top to bottom, the left panels show the calibrated ZTF PSF light curves of the target AGN and the two comparison stars (S1 and S2), followed by the variation of the PSF FWHM during the observing session. The corresponding right panels show the median-subtracted star--star ($S1-S2$) DLC and the two target--star ($AGN-S1$ and $AGN-S2$) DLCs derived from our aperture photometry, followed by the PSF FWHM. The same pair of comparison stars and the same observing session are used in both analyses. The individual ZTF light curves also allow the stability of the two comparison stars to be directly assessed.
    }
    \label{fig:j0731_comparison}
\end{figure}

% J0757
\begin{figure}
    \centering
    \includegraphics[width=0.95\textwidth]{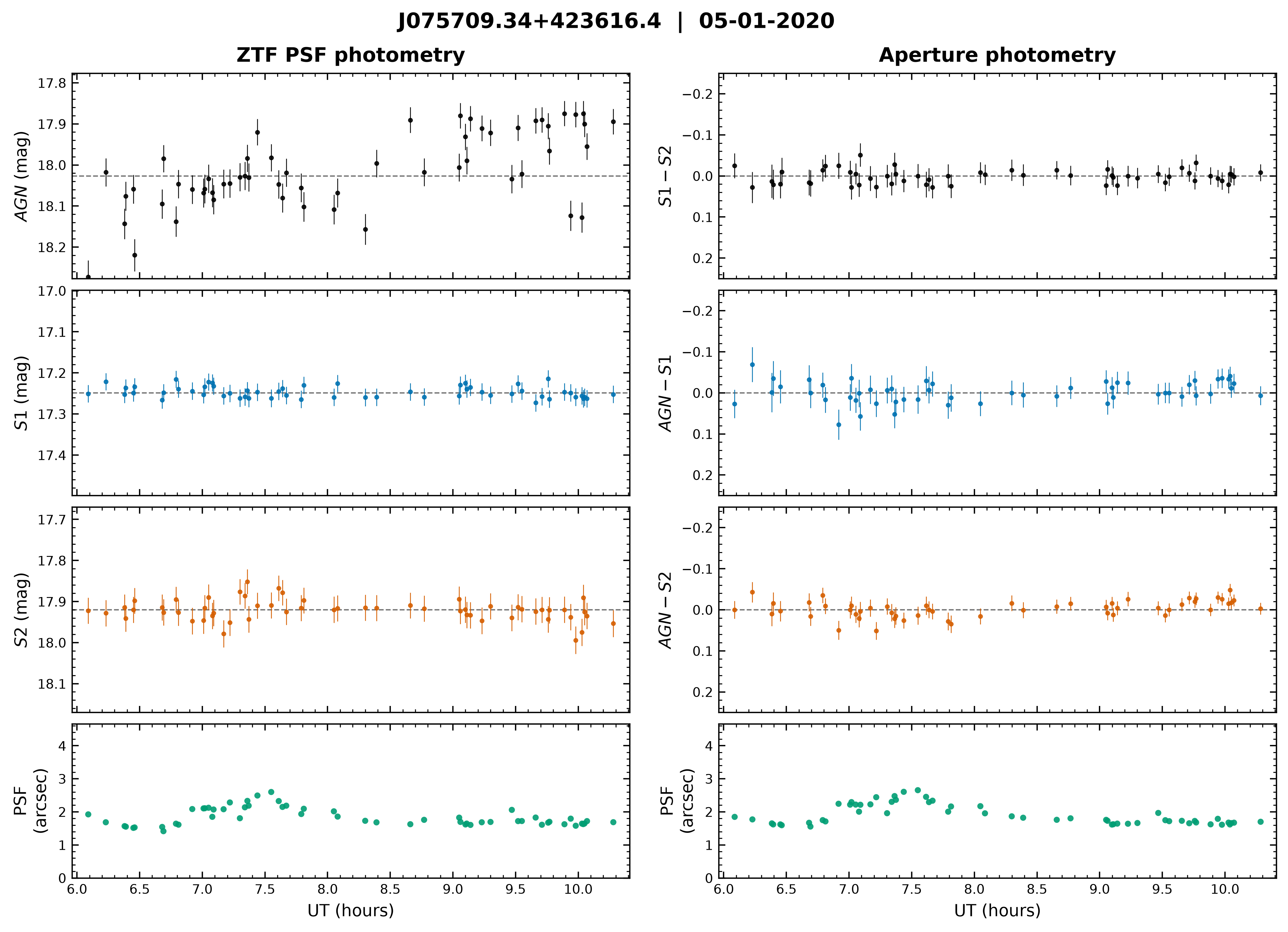}
    \caption{
    Same as Fig.~\ref{fig:j0731_comparison}, but for the LMAGN J075709.34+423616.4, observed on 5 January 2020. The left panels show the calibrated ZTF PSF light curves of the AGN, S1, and S2 individually, while the right panels show the corresponding DLCs obtained from our aperture photometry.
    }
    \label{fig:j0757_comparison}
\end{figure}

% J0815
\begin{figure}
    \centering
    \includegraphics[width=0.95\textwidth]{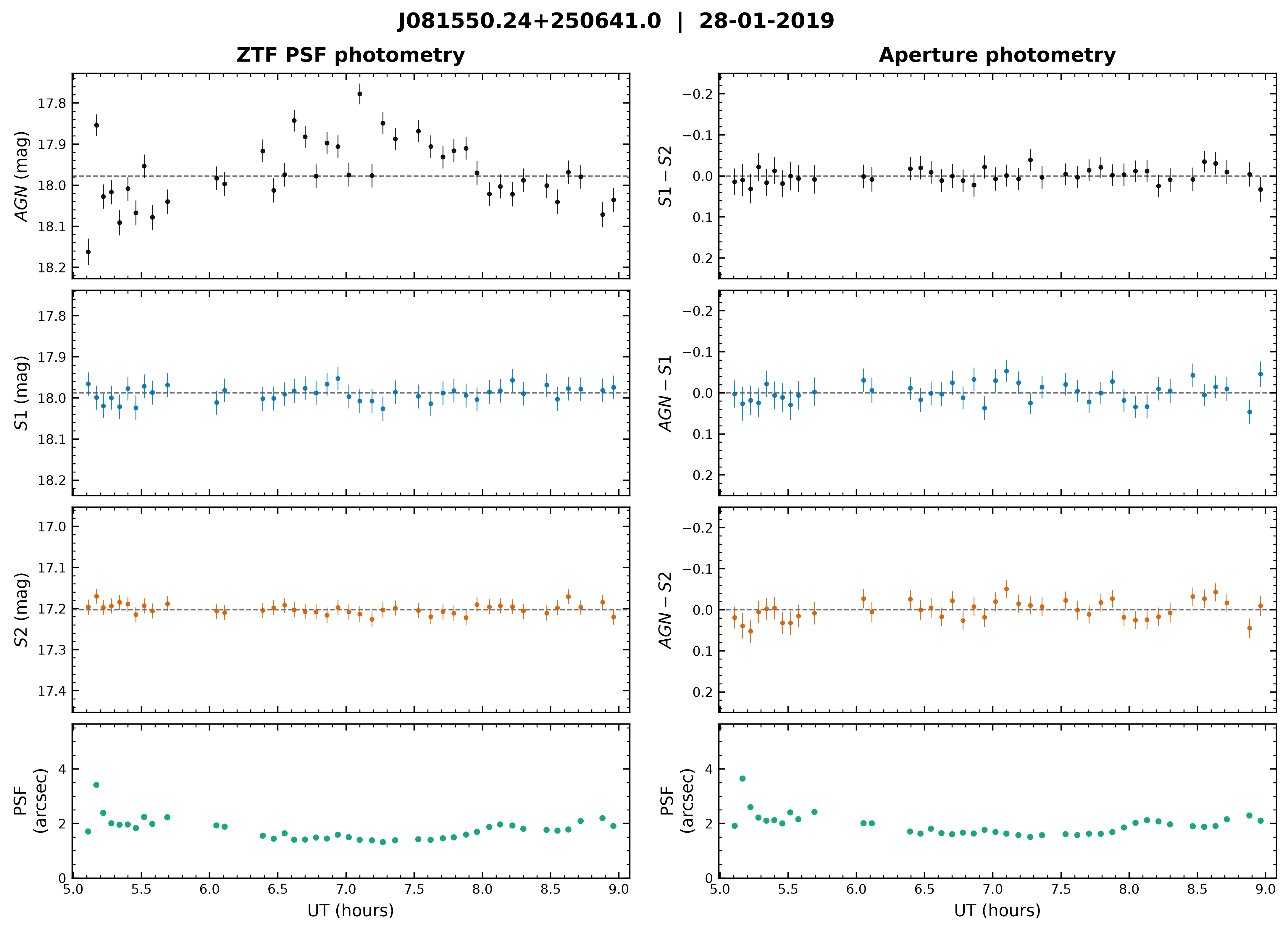}
    \caption{
    Same as Fig.~\ref{fig:j0731_comparison}, but for the LMAGN J081550.24+250641.0, observed on 28 January 2019. The left panels show the calibrated ZTF PSF light curves of the AGN, S1, and S2 individually, while the right panels show the corresponding DLCs obtained from our aperture photometry.
    }
    \label{fig:j0815_comparison}
\end{figure}

% J0830
\begin{figure}
    \centering
    \includegraphics[width=0.95\textwidth]{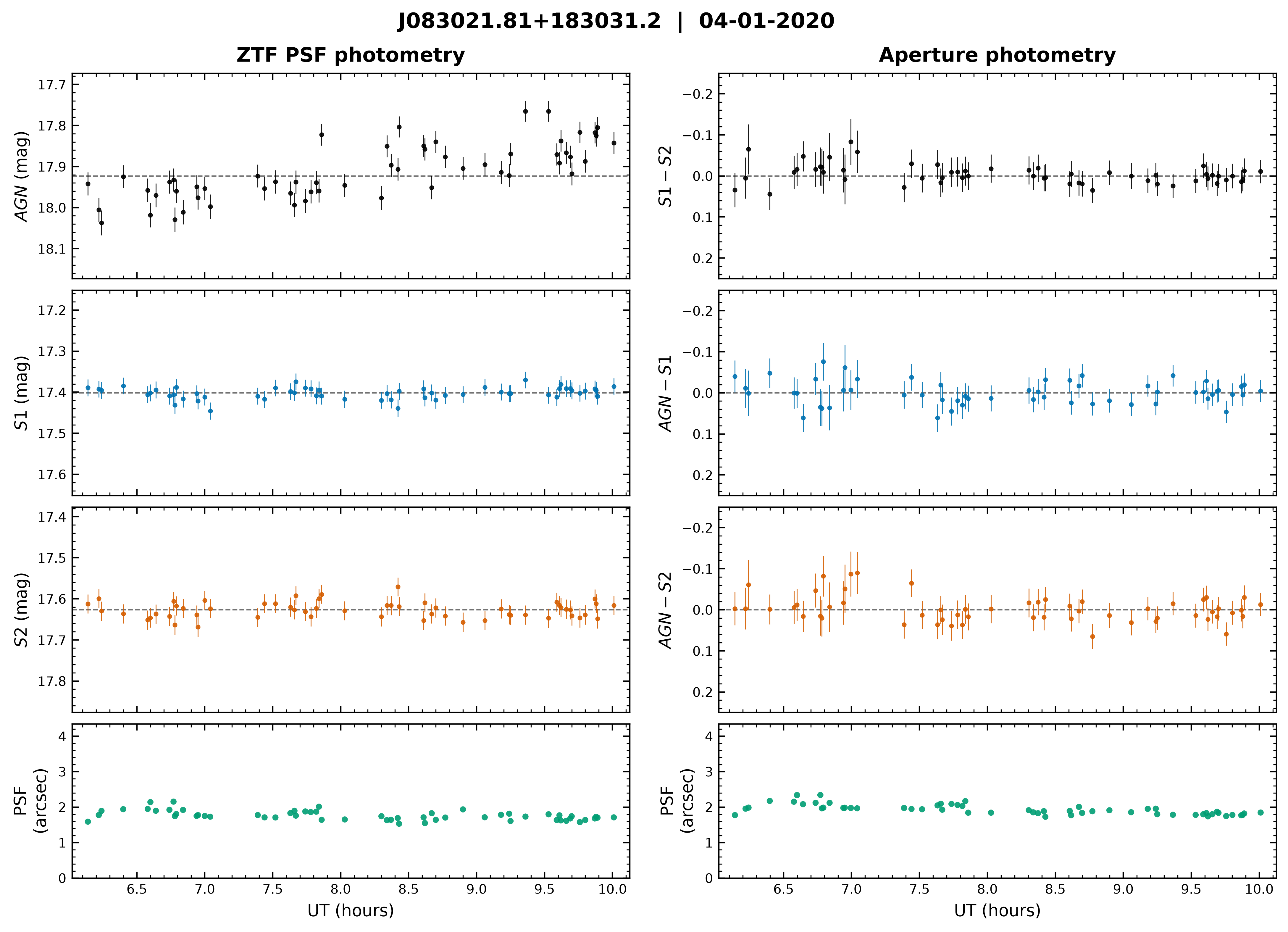}
    \caption{
    Same as Fig.~\ref{fig:j0731_comparison}, but for the LMAGN J083021.81+183031.2, observed on 4 January 2020. The left panels show the calibrated ZTF PSF light curves of the AGN, S1, and S2 individually, while the right panels show the corresponding DLCs obtained from our aperture photometry.
    }
    \label{fig:j0830_comparison}
\end{figure}

\bsp
\label{lastpage}
\end{document}